\documentclass[11pt]{article}

\usepackage[utf8]{inputenc}
\usepackage[T1]{fontenc}

\usepackage{amsmath,amssymb,amsthm,bbm}

\usepackage{graphicx,transparent,eso-pic}

\usepackage[document]{ragged2e}
\usepackage{pagecolor,color}
\usepackage{soul}

\usepackage{ifthen}

\usepackage{natbib}

\newcommand{\noun}[1]{\textsc{#1}}

\usepackage{url}

\usepackage[margin=2cm]{geometry}

\begin{document}

\title{Modeling the co-evolution of cities and networks}
\author{\noun{Juste Raimbault}$^{1,2}$\\
$^1$ UPS CNRS 3611 ISC-PIF\\
$^2$ UMR CNRS 8504 G{\'e}ographie-cit{\'e}s
}
\date{}

\maketitle

\justify

\begin{abstract}
The complexity of interactions between networks and territories has been widely acknowledged empirically, in particular through the existence of circular causal relations in their co-development, that can be understood as a co-evolution. This contribution aims at investigating models that endogenize this co-evolution, in the particular case of cities and transportation networks. We introduce a family of models of co-evolution for systems of cities at the macroscopic scale. Interactions between cities are the main driver of population growth rates, capturing a network effect at the first order (direct interactions). Network growth follows a demand-induced thresholded growth scheme, that can occur at the global level or locally. The exploration of the model on synthetic systems of cities shows the ability of the model to capture co-evolutive patterns. We apply the model on the French system of cities, with population data spanning 1831-1999 and a dynamical railway network (1850-2000). The model is calibrated on successive time-windows, assuming local temporal stationarity. We extract therein indirect knowledge on underlying processes and find that the prediction for city populations are in some cases improved in comparison to a static model.
\end{abstract}


\section{Introduction}

\subsection{Structuring effects of transportation networks}

The idea of possible causal relationships between territorial characteristics and transportation networks has fed a scientific debate that is still active nowadays. The underlying assumptions can be synthesized as more or less deterministic attributions of impacts of transportation infrastructures or of a new transportation mode on territorial transformations. Precursors of such a reasoning can be tracked back in the twenties: \cite{burgess1925city} mentions for example some ``modifications of forms of transportation and communication as determining factors of growth and decline cycles [of territories]'' (p.~69). Methodologies to identify what is then called \emph{structuring effects} of transportation networks have been developed for planning in the seventies: \cite{bonnafous1974methodologies} situate the concept of structuring effects in the perspective of using the transportation offer as a planning tool (the alternatives are the development of an offer to answer to a congestion of the network, and the simultaneous development of associated offer and planning). These authors identify from an empirical viewpoint direct effects of a novel offer on the behavior of agents, on transportation flows and possible inflexions on socio-economic trajectories of concerned territories. \cite{bonnafous1974detection} develop a method to identify such effects through the modification of the class of cities in a typology established a posteriori. More recently, \cite{bonnafous2014observatoires} recalls that the institution of \emph{permanent observatories} for territories makes such analyses more robust, allowing a continuous monitoring of the territories that are the most concerned by the extent of a new infrastructure.

According to \cite{offner1993effets} which reformulates ideas already given by~\cite{franccois1977autoroutes} for example, a not reasoned and out-of-context use of these methods has then been developed by planners and politicians which generally used them to justify transportation projects in a technocratic manner: through the argument of a direct effect of a new infrastructure on local development (for example economic), politics are able to ask for subsidies and to legitimate their action in front of the people. \cite{offner1993effets} insists on the necessity of a critical positioning on these issues, recalling that there exists no scientific demonstration of an effect that would be systematic. A special issue of the journal \emph{L'Espace Géographique}~\citep{espacegeo2014effets} on that debate recalled that on the one hand misconceptions and misuses were still greatly present in operational and planning communities, which can be explained for example by the need to justify public actions, and on the other hand that a scientific understanding of relations between networks and territories is still in construction.



\subsection{Co-evolution of cities and networks}

An alternative approach to relations between transportation networks and territories is to consider them as \emph{co-evolving}. The evolutive urban theory considers systems of cities as systems of systems at multiple scale, from the intra-urban microscopic level, to the macroscopic level of the whole system, through the mesoscopic level of the city~\citep{pumain2008socio}. These systems are complex, dynamical, and adaptive: their components \emph{co-evolve} and the system answers to internal or external perturbations by modifying its structure and its dynamics. Interactions between cities are the main drivers of these evolutionary patterns. These interactions consist in material or informational exchanges, and the diffusion of innovation is therein a crucial component~\citep{pumain2010theorie}. These are necessarily carried by physical networks, and more particularly transportation networks. We expect thus from a theoretical point of view strong interdependencies between cities and transportation networks at these scales, i.e. a co-evolution.

From the empirical point of view, it has already been suggested by some studies: \cite{bretagnolle2003vitesse} reveals an increasing correlation in time between urban hierarchy and the hierarchy of temporal accessibility for the French railway network (which is a priori clearer for this measure than for integrated measures of accessibility that are prone to auto-correlation). This correlation is a clue of positive feedbacks between urban ranks and network centralities. According to \cite{bretagnolle:tel-00459720}, different regimes of interactions between cities and transportation networks have been identified: for the evolution of the French railway network, a first phase of adaptation of the network to the existing urban configuration was followed by a phase of co-evolution, in the sense that causal relations became difficult to identify. The impact of the contraction of space-time by networks on patterns of growth potential had already been shown for Europe with an exploratory analysis in~\citep{bretagnolle1998space}.

Modeling results by~\cite{bretagnolle2010comparer}, and more particularly the different parametrizations of the Simpop2 model, unveil different regimes of interactions between networks and cities. The generic structure of the Simpop2 model is the following~\citep{pumain2008socio}: cities are characterized by their population ad their wealth; they product goods according to their economic profile; interactions between cities produce exchanges, determined by the offer and demand functions; populations evolve according to wealth after exchanges. The application of this model shows that the evolution of the railway network in the United States has followed a rather different dynamic, without hierarchical diffusion, shaping locally urban growth in some cases. This particular context of conquest of a space empty of infrastructures implies a specific regime for the territorial system. Other contexts reveal different impacts of the network at short and long term: \cite{berger2017locomotives} study the impact of the construction of the Swedish railway network on the growth of urban populations, from 1800 to 2010, and find an immediate causal effect of the accessibility increase on population growth, followed on long times of a strong inertia for population hierarchy. In each case, we indeed observe the existence of \emph{structural dynamics} on long times, which correspond to the slow dynamics of the urban system structure, and witness in that sense of \emph{structuring effects on long times} as~\cite{pumain2014effets} puts it.

We however differentiate the latest from the structuring effects previously mentioned which are subject to debates. At the level of the urban system, it is relevant to globally follow trajectories that were possible, and locally the effect has necessarily a probabilistic aspect. Moreover, we insist on the role of path-dependency for trajectories of urban systems: for example the existence in France of a previous system of cities and network (postal roads) has strongly influenced the development of the railway network, or as \cite{berger2017locomotives} showed for Sweden. The same way, \cite{doi:10.1068/b39089} highlight the importance of historical events in coupled dynamics of the road network and territories, historical shocks that can be seen as exogenous and inducing bifurcations of the system that accentuate the effect of path-dependency. Therefore, for these structural dynamics on long times, forecasting can difficultly be considered.

\subsection{Models of co-evolution}

Considering models as a fundamental component for the production of knowledge, in the sense of a knowledge domain in itself \citep{raimbault2017applied}, this chapter will focus on simulation models concerned with the co-evolution of cities and networks. We now propose to review models that integrate dynamically a strong coupling between territorial components and transportation networks, in the sense that a clear conditioning of one by the other can not be identified. We will broadly designate by model of co-evolution simulation models that include a coupling of urban growth dynamics and transportation network growth dynamics. These are relatively rare, and for most of them still at the stage of stylized models. The efforts being relatively sparse and in very different domains, there is not much unity in these approaches, beside the abstraction of the assumption of an interdependency between networks and territorial characteristics in time. This sparsity may be due to a high compartmentalization of related disciplines \citep{raimbault2017models}. We propose now to review them through the prism of scales.


\subsubsection{Microscopic and mesoscopic scales}

\paragraph{Geometrical Models}

\cite{achibet2014model} describes a co-evolution model at a very large scale (scale of the building), in which evolution of both network and buildings are ruled by a same agent, influenced differently by network topology and population density, and that can be understood as an agent of urban development. The model allows to simulate an auto-organized urban extension and to produce district configurations. Even if it strongly couples territorial components (buildings) and the road network, described results do not imply any conclusion on the processes of co-evolution themselves.

A generalization of a geometrical local optimization model for network growth yield a co-evolution model for network topology and the density of its nodes \citep{barthelemy2009co}. The localization of new nodes is simultaneously influenced by density and centrality, yielding the looping of the strong coupling. More precisely, the global behavior of the model is the same, as the network extension behavior. Centers then localize following a utility function that is a linear combination of average betweenness centrality in a neighborhood and of the opposite of density (dispersion due to higher price as a function of density). This utility is used to compute the probability of localization of new centers following a discrete choices model. The model allows to show that the influence of centrality reinforces aggregation phenomena (in particular through an analytical resolution on a one-dimensional version of the model), and furthermore reproduces exponentially decreasing density profiles (Clarcke's law) which are observed empirically.

\cite{ding2017heuristic} introduce a model of co-evolution between different layers of the transportation network, and show the existence of an optimal coupling parameter in terms of inequalities for the centrality in network conception: if the road network is assimilated at a fine granularity to a population distribution, this model can be compared with the precedent model of co-evolution between the transportation network and the territory.

\paragraph{Economic models}

\cite{levinson2007co} take an economic approach, which is richer from the point of view of network development processes implied, similar to a four step model (i.e. including the generation of origin-destination flows and the assignment of traffic in the network) which takes into account travel cost and congestion, coupled with a road investment module simulating toll revenues for constructing agents, and a land-use evolution module updating actives and employments through discrete choice modeling. The exploration experiments show that co-evolving network and land uses lead to positive feedbacks reinforcing hierarchies. These are however far from satisfying, since network topology does not evolve as only capacities and flows change within the network, what implies that more complex mechanisms (such as the planning of new infrastructures) on longer time scales are not taken into account. \cite{li2016integrated} have recently extended this model by adding endogenous real estate prices and an optimization heuristic with a genetic algorithm for deciding agents.

From an other point of view, \citep{levinson2005paving} is also presented as a model of co-evolution, but corresponds more to a predictive model based on Markov chains, and thus closer to a statistical analysis than a simulation model based on these processes. \cite{rui2011urban} describe a model in which the coupling between land-use and network topology is done with a weak paradigm, land-use and accessibility having no feedback on network topology, the land-use model being conditioned to the growth of the autonomous network.

\paragraph{Cellular automatons}

A simple hybrid model explored and applied to a stylized planning example of the functionnal distribution of a new district in~\citep{raimbault2014hybrid}, relies on mechanisms of accessibility to urban activities for the growth of settlements with a network adapting to the urban shape. The rules for network growth are too simple to capture more elaborated processes than just a simple systematic connection (such as potential breakdown for example), but the model produces at a large scale a broad range of urban shapes reproducing typical patterns of human settlements. This model is inspired by~\citep{moreno2012automate} for its core mechanisms but yield a much broader generation of forms by taking into account urban functions.

At these relatively large scales, spanning from the urban to the metropolitan scale, mechanisms of population localization influenced by accessibility coupled to mechanisms of network growth optimizing some particular functions seem to be the rule for this kind of models: in the same way, \cite{wu2017city} couple a cellular automaton for population diffusion to a network optimizing local cost that depends on the geometry and on population distribution.

Models answering to more remote questions can furthermore be linked to our problem: for example, in a conceptual way, a certain form of strong coupling is also used in \citep{bigotte2010integrated} which by an approach of operational research propose a network design algorithm to optimize the accessibility to amenities, taking into account both network hierarchy and the hierarchy of connected centers.

This way, co-evolution models at the microscopic and mesoscopic scales globally have the following structure: (i) processes of localization or relocalization of activities (actives, buildings) influenced by their own distribution and network characteristics; (ii) network evolution, that can be topological or not, answering to very diverse rules: local optimization, fixed rules, planning by deciding agents. This diversity suggests the necessity to take into account the superposition of multiple processes ruling network evolution.

\subsubsection{Urban systems modeling}

At a macroscopic scale, co-evolution can be taken into account in models of urban systems. \cite{baptiste1999interactions} propose to couple an urban growth model based on migrations (introduced by the application of synergetics to systems of cities by~\citep{sanders1992systeme}) with a mechanism of self-reinforcement of capacities for the road network without topological modification. More precisely, the general principles of the model are the following: (i) attractivity and repulsion indicators allow for each city to determine emigration and immigration rates and to make populations evolve; (ii) network topology is fixed in time, but capacities of links evolve. The rule is an increase in capacity when the flow becomes greater given a fixed parameter threshold during a given number of iterations. Flows are affected with a gravity model of interaction between cities. The last version of this model is presented by~\cite{baptistemodeling}. General conclusions that can be obtained from this work are that this coupling yield a hierarchical configuration and that the addition of the network produces a less hierarchical space, allowing medium-sized cities to benefit from the feedback of the transportation network. These conclusions remain limited as simpler models without co-evolution such as the one developed by \cite{raimbault2018indirect} also produce hierarchical urban systems.

The model proposed by~\cite{blumenfeld2010network} can be seen as a bridge between the mesoscopic scale and the approaches of urban systems, since it simulates migrations between cities and network growth induced by potential breakdown when detours are too large. In the continuity of Simpop models for systems of cities, \cite{schmitt2014modelisation} describes the SimpopNet model which aims at precisely integrating co-evolution processes in systems of cities on long time scales, typically via rules for hierarchical network development as a function of the dynamics of cities, coupled with these that depends on network topology. Unfortunately the model was not explored nor further studied, and furthermore stayed at a toy-level. \cite{cottineau2014evolution} proposes an endogenous transportation network growth as the last building brick of the Marius modeling framework, but it stays at a conceptual level since this brick has not been specified nor implemented yet. To the best of our knowledge, there exists no model which is empirical or applied to a concrete case based on an approach of co-evolution by urban systems from the point of view of the evolutive urban theory.

We must note here the epistemological opposition of these geographical approaches inspired by the evolutive urban theory \citep{pumain1997pour} to principles of economic geography: \cite{fujita1999evolution} introduce for example an evolutionary model able to reproduce and urban hierarchy and an organization typical of central place theory~\citep{banos2011christaller}, but that still relies on the notion of successive equilibriums, and moreover considers a ``Krugman-like'' model, i.e. a one dimensional and isotropic space, in which agents are homogeneously distributed. This approach can be instructive on economic processes in themselves but more difficultly on geographical processes, since these imply the embedding of economic processes in the geographical space which spatial particularities not taken into account in this approach are crucial. Our work will focus on demonstrating to what extent this structure of space can be important and also explicative, since networks, and even more physical networks induce spatio-temporal processes that are path-dependent and thus sensitive to local singularities and prone to bifurcations induced by the combination of these with processes at other scales (for example the centrality inducing a flow).

We can mention several other streams of research which are concerned with modeling co-evolution in systems of cities, such as economic geography \citep{schamp201020}. For example, \cite{doi:10.1080/00343400802662658} models the co-evolution of firms and networks. \cite{liu2013exploring} empirically study the co-evolution of air transportation networks and companies networks. \cite{neal2012structural} uses network models to investigate the concept of interlocking of firms and cities, of which dynamics can be understood as co-evolutive. Closer to the fields of planning and transportation, Land-use Transport Interaction (LUTI) models do not explicitly include co-evolution but are concerned with coupling aspects of both \citep{wegener2004land}. These class of models can be generalized at the macroscopic scale to model larger systems of cities \citep{russo2012unifying}.

To synthesize the approaches at the macroscopic scale, existing models are mostly based on the evolution of agents (generally cities) as a consequence of their interactions, carried by the network, whereas the evolution of the network can follow different rules: self-reinforcement, potential breakdown. The general structure is globally the same than at larger scales, but ontologies are fundamentally different.

\subsection{Proposed approach}

This literature review confirms the rarity of co-evolution models for cities and transportation networks at the macroscopic scale of a system of cities. Concerning the existing examples, \cite{baptistemodeling} is too specific on the processes included and the case study to which it is applied, whereas \cite{schmitt2014modelisation} does not include other network growth processes than topological breakdown and exogenously conditions link speeds. A generic family of models that can both apply to stylized and real systems of cities, and with a certain flexibility in processes included, would therefore fill this gap in the literature.

We propose in this chapter to introduce such a family of co-evolution models. Our contribution is significant on several points: (i) this is to the best of our knowledge that such a generic model is introduced to model the co-evolution of cities and transportation networks; (ii) we systematically explore a specification of the model on synthetic systems of cities and unveil its ability to produce co-evolutive regimes; (iii) we calibrate it on the French system of cities with population and railway network data.

The rest of this chapter is organized as follows: we first formalize the family of models and the specification studied in the following; we then systematically explore its behavior on synthetic systems of cities, and describe its calibration for the French system of cities. We finally briefly present an alternative specification for network growth and discuss possible developments and applications of this work.

\section{A family of co-evolution models}

We introduce in this section the family of models we propose for the co-evolution of cities and networks at a macroscopic scale.

\subsection{Rationale}

Our approach relies in a direct extension of the interaction model within a system of cities described by \cite{raimbault2018indirect}, at a macroscopic scale with an ontology typical to systems of cities. For the sake of simplicity, we similarly stick to an unidimensional description of cities by their population.

Concerning network growth, we propose also to stay at a relatively aggregated and simplified level, allowing to test growth heuristics at different levels of abstraction. In order to be flexible on model mechanisms, diverse processes can be taken into account, such as direct interactions between cities, intermediate interactions through the network, the feedback of network flows and a demand-induced network growth. Empirical characteristics emphasized by~\cite{thevenin2013mapping} for the French railway network suggest the existence of feedbacks of network use, or of flows traversing it, on its persistence and its development, whose properties have evolved in time: a first phase of strong development would correspond to an answer to a high need of coverage, followed by a reinforcement of main links and the disappearance of weakest links.

\begin{figure}
\includegraphics[width=\linewidth]{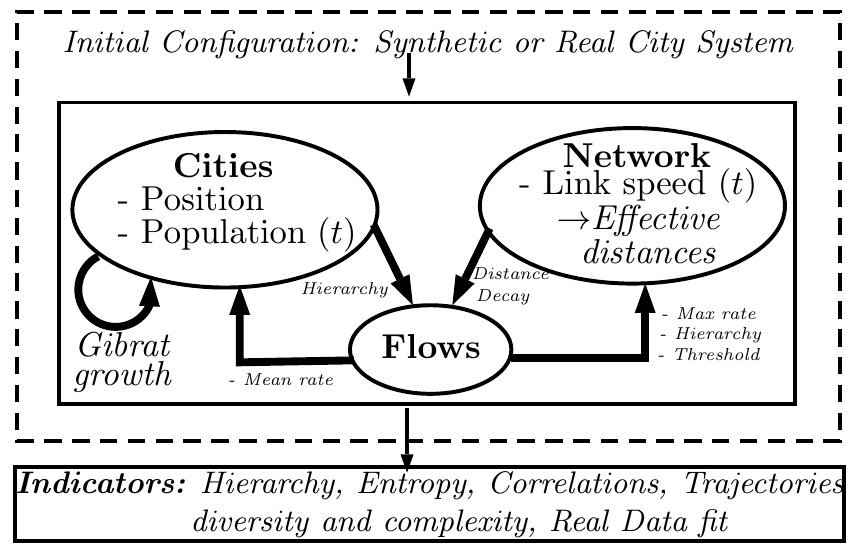}
\caption{\textbf{Abstract representation of the model.} Ellipses correspond to main ontological elements (cities, network, flows), whereas arrows translate processes for which associated parameters are given. The model is described in its broader ecosystem of initialisation and output indicators.\label{fig:macrocoevol:model}}
\end{figure}

The coupling between cities and the network is achieved by the intermediate of flows between cities in the network: these capture the interactions between cities and have simultaneously an influence on the network in which they flow. The Fig.~\ref{fig:macrocoevol:model} shows the structure of the model in terms of geographical objects and processes included and their relations.

\subsection{Model description}

\subsubsection{Generic formulation}

The urban system is characterized by populations $\mu_i(t)$ and the network $\mathbf{G}(t)$, to which can be associated a distance matrix $d^G_{ij}(t)$. Flows between cities $V_{ij}$ are given by a gravity interaction which writes

\begin{equation}
V_{ij} = \left(\frac{\mu_i\mu_j}{\left(\sum_k{\mu_k}\right)^2}\right)^{\gamma_G}\cdot \exp{\left(-d_{ij}/d_G\right)}
\end{equation}

with $d_ij$ network distance.

The evolution of populations then follows the specifications of \cite{raimbault2018indirect}, i.e.

\begin{equation}
\vec{\mu}(t+1)=\Delta t\cdot \left[ r_0\cdot \mathbf{Id}\cdot \vec{\mu} + w_G\cdot \sum_j \frac{V_{ij}}{<V_{ij}>}\right]
\end{equation}

Concerning the network, we assume that it evolves following the equation
\begin{equation}
\mathbf{G}(t + 1) = F(\mathbf{G}(t),\phi_{ij}(t))
\end{equation}
such that the assignment of flows within the network and a local variation of its elements can be taken into account through the function $F$.

This most generic form describes our family of models, which can yield different specifications of models, in particular by describing the network growth heuristic. Note that this model could be easily extended similarly to \cite{raimbault2018indirect}, by adding second-order effects of network centralities on population growth for example.

\subsubsection{Distance-based network growth specification}

We propose in a first time to consider patterns linked to distance only, and to specify a relation on an abstract network as
\begin{equation}
d^G_{ij}(t+1) = F(d^G_{ij}(t),\phi_{ij}(t))
\end{equation}
i.e. an evolution of the distance matrix only. In this spirit, we keep an interaction model strictly at a macroscopic scale, since a precise spatialization of the network would imply to take into account a finer scale that includes the local shape of the network which determines shortest paths.

Following a thresholded feedback heuristic, given a flow $\phi$ in a link, we assume its effective distance to be updated by:

\begin{equation}
d(t+1) = d(t)\cdot \left( 1 + g_{max} \cdot \left[\frac{1 - \left(\frac{\phi}{\phi_0}\right)^{\gamma_s}}{1 + \left(\frac{\phi}{\phi_0}\right)^{\gamma_s}}\right]\right)
\end{equation}

with $\gamma_s$ a hierarchy parameter, $\phi_0$ the threshold parameter and $g_{max}$ the maximal growth rate at each step. This auto-reinforcement function can be interpreted the following way: above a limit flow $\phi_0$, the travel conditions improve, whereas they deteriorate below. The hierarchy of gain is given by $\gamma_s$, and since $\frac{1 - \left(\frac{\phi}{\phi_0}\right)^{\gamma_s}}{1 + \left(\frac{\phi}{\phi_0}\right)^{\gamma_s}} \rightarrow_{\phi\rightarrow \infty} -1$, $g_{max}$ is the maximal distance gain. This function is similar to the one proposed by \cite{tero2007mathematical}, which uses $\Delta d = \Delta t \left[ \frac{\phi^\gamma}{1 + \phi^\gamma} - d\right]$. This function yield similarly a threshold effect, since the derivative vanishes at $\phi^{\ast} = \left(\frac{d}{1 - d}\right)^{1/\gamma}$, but this value depends on the distance and can more difficultly be adjusted to a value that can be interpreted.

Our specification is summarized in Fig.~\ref{fig:macrocoevol:model} as a model workflow. We indeed have a double feedback of populations and network on themselves, but also the interplay of one on the other, through the mediation of network flows. As interaction potentials depends on populations and network at a given time, the full history of both is crucial to determine it, and the model takes thus into account path-dependancy of the system.

\subsection{Indicators}

To quantify the behavior of the model, we use different indicators, that give a grasp on typical properties of systems of cities produced by the model and on typical properties of temporal trajectories. These have been introduced by \cite{2018arXiv180900861R} precisely to study the behavior of a macroscopic model of co-evolution for cities and transportation networks, in a very similar context.
\begin{itemize}
	\item summary statistics of trajectories, taken as hierarchy $\alpha\left[\cdot\right]$, entropy $\varepsilon\left[\cdot\right]$ and average $\bar{\cdot}$ of distributions;
	\item rank correlations $\rho\left[\cdot\right]$ between initial state and final state;
	\item diversity $D\left[\cdot\right]$ and complexity $C\left[\cdot\right]$ of temporal trajectories.
\end{itemize}

These indicators are applied on variables representing both cities and the transportation network: populations $\mu_i(t)$, closeness centralities $c_i (t)$ and generalized accessibilities $Z_i (t)$.

\subsection{Parameter space}

The parameter space of the specific model is composed of 7 parameters, namely $r_0$ the endogenous growth rate, $w_G$ the weight of interactions, $d_G$ the spatial range of interactions, $\gamma_G$ the hierarchy of interactions, $g_{max}$ the maximal network growth, $\phi_0$ the network threshold and $\gamma_S$ the network growth hierarchy. We summarize in Table~\ref{tab:parameters} these parameters with the corresponding processes and theoretical ranges.

\begin{table}
\caption{\textbf{Parameters of the model.} We give the parameters of the model that vary in experiments, with the corresponding processes, interpretation and theoretical ranges.\label{tab:parameters}}	
\begin{tabular}{|l|l|l|l|l|}
\hline
Parameter & Notation & Process & Interpretation & Range\\
\hline
Growth rate & $r_0$ & Endogenous growth & Urban growth & $\left[ 0,1\right]$ \\
Gravity weight & $w_G$ & Direct interaction & Maximal growth & $\left[ 0,1\right]$ \\
Gravity gamma & $\gamma_G$ & Direct interaction & Level of hierarchy & $\left[ 0,+\infty\right]$ \\
Gravity decay & $d_G$ & Direct interaction & Interaction range & $\left[ 0,+\infty\right]$ \\
Maximal network growth & $g_{max}$ & Network growth & Speed improvement & $\left[ 0,+\infty\right]$ \\
Network threshold & $\phi_0$ & Network growth & Use threshold & $\left[ 0,+\infty\right]$ \\
Network growth hierarchy & $\gamma_S$ & Network growth & Reinforcement strength & $\left[ 0,+\infty\right]$ \\
\hline
\end{tabular}
\end{table}

\section{Results}



The model is fully implemented in NetLogo, for the simplicity of coupling between heterogeneous components. A particular care is taken for the duality of network representation, both as a distance matrix and as a physical network, in order to facilitate the extension to physical network heuristics. Source code, data and results are openly available on the git repository of the project at \url{https://github.com/JusteRaimbault/CityNetwork/tree/master/Models/MacroCoevol}. Simulation results are available on the dataverse data repository at \url{http://dx.doi.org/10.7910/DVN/TYBNFQ}.

The model was explored and calibrated using the OpenMole workflow engine \citep{reuillon2013openmole}, which provide exploration methods and allows a transparent distribution of computation jobs on a computation grid.

\subsection{Exploration on a synthetic system of cities}

\subsubsection{Synthetic setup}

The model is first tested and explored on synthetic systems of cities as it is done by \cite{favaro2011gibrat}, in order to understand some of its intrinsic properties. In this case, we consider the model with an abstract network as specified above, i.e. without spatial description of the network and with evolution rules acting directly on $d^G_{ij}$ given the previous specifications. A synthetic city system is generated following the heuristic used in the previous section: (i) $N_S$ cities are randomly distributed in a homogeneous geographical space; (ii) populations are attributed to cities following an inverse power law, with a hierarchy parameter $\alpha_S$ and such that the largest city has a population equal to $P_{max}$, i.e. following $P_i = P_{max} \cdot i^{-\alpha_S}$.

To simplify, several meta-parameters are fixed: the number of cities is fixed at $N_S = 30$, the maximal population at $P_{max} = 100000$ and the maximal network growth to $g_{max} = 0.005$. Final time is fixed at $t_f = 30$, what corresponds to distances divided approximatively by 5, in order to comply to an empirical constraint: this corresponds to the evolution of the travel time between Paris and Lyon from around ten hours at the beginning of the century to two hours today, showed for example by~\cite{thevenin2013mapping}. We also neglect network effects at the second order by taking $w_N = 0$. We explore a grid in the extended parameter space $\alpha_S$, $\phi_0$, $\gamma_s$, $w_G$, $d_G$, $\gamma_G$. We describe the results for $\alpha_S = 1$, what is the closest to existing city systems (in comparison to 0.5 and 1.5, see the systematic review of the rank-size law estimations done by~\cite{10.1371/journal.pone.0183919}).

\subsubsection{Model behavior}

\begin{figure}
\includegraphics[width=\linewidth,height=0.9\textheight]{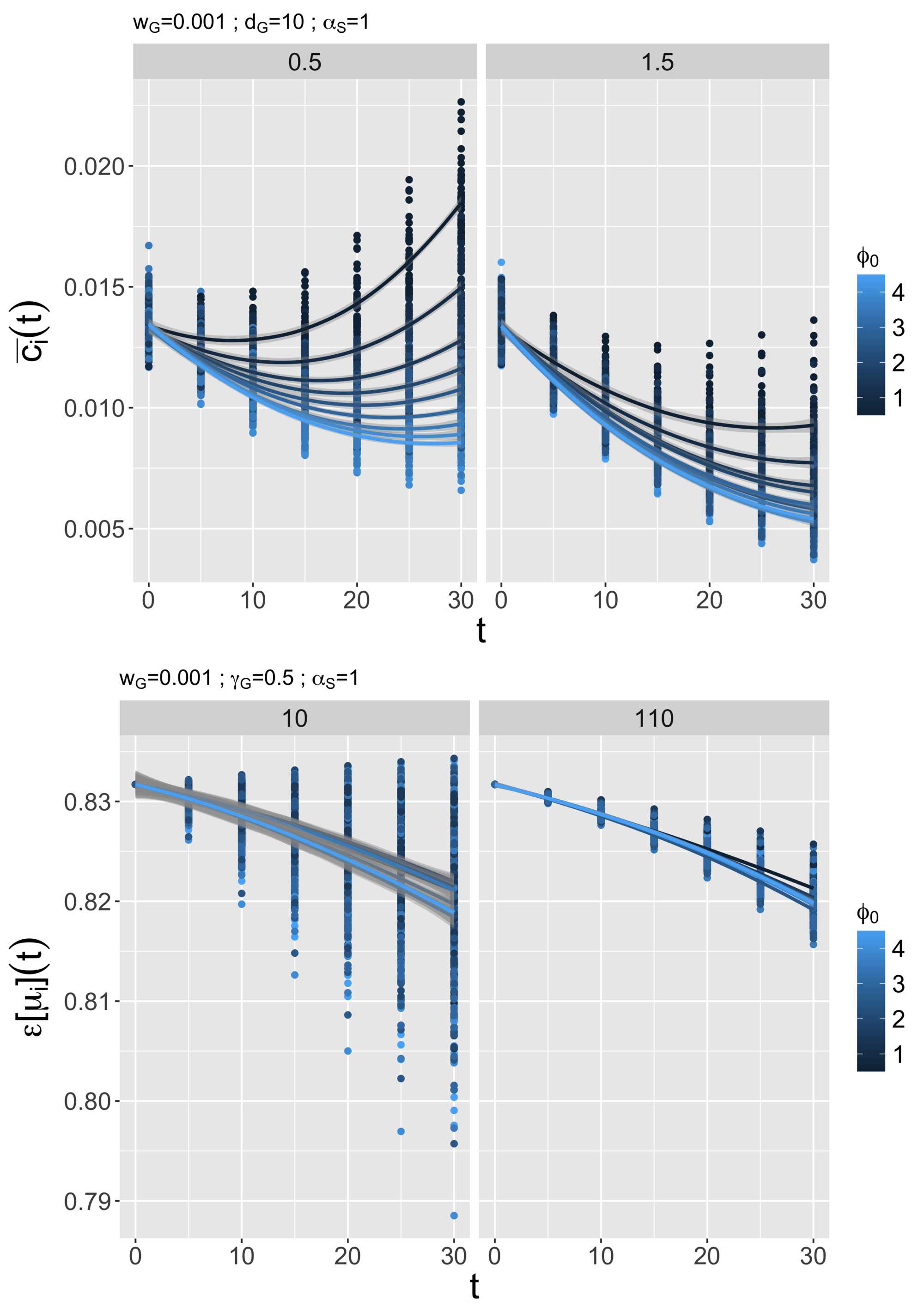}
\caption{\textbf{Temporal behavior of the co-evolution model with abstract network on a synthetic system of cities.} \textit{(Top)} Average closeness centralities, as a function of time, for $\gamma_G$ (rows) and $\phi_0$ (color) variable, at fixed $w_G = 0.001$ and $d_G = 10$; \textit{(Bottom)} Entropy of populations, as a function of time, for $d_G$ (columns) and $\phi_0$ (color) variable, at fixed $w_G = 0.001$ and $\gamma_G = 0.5$. See main text for interpretation.\label{fig:macrocoevol:behavior-time}}
\end{figure}

\begin{figure}
\includegraphics[width=\linewidth]{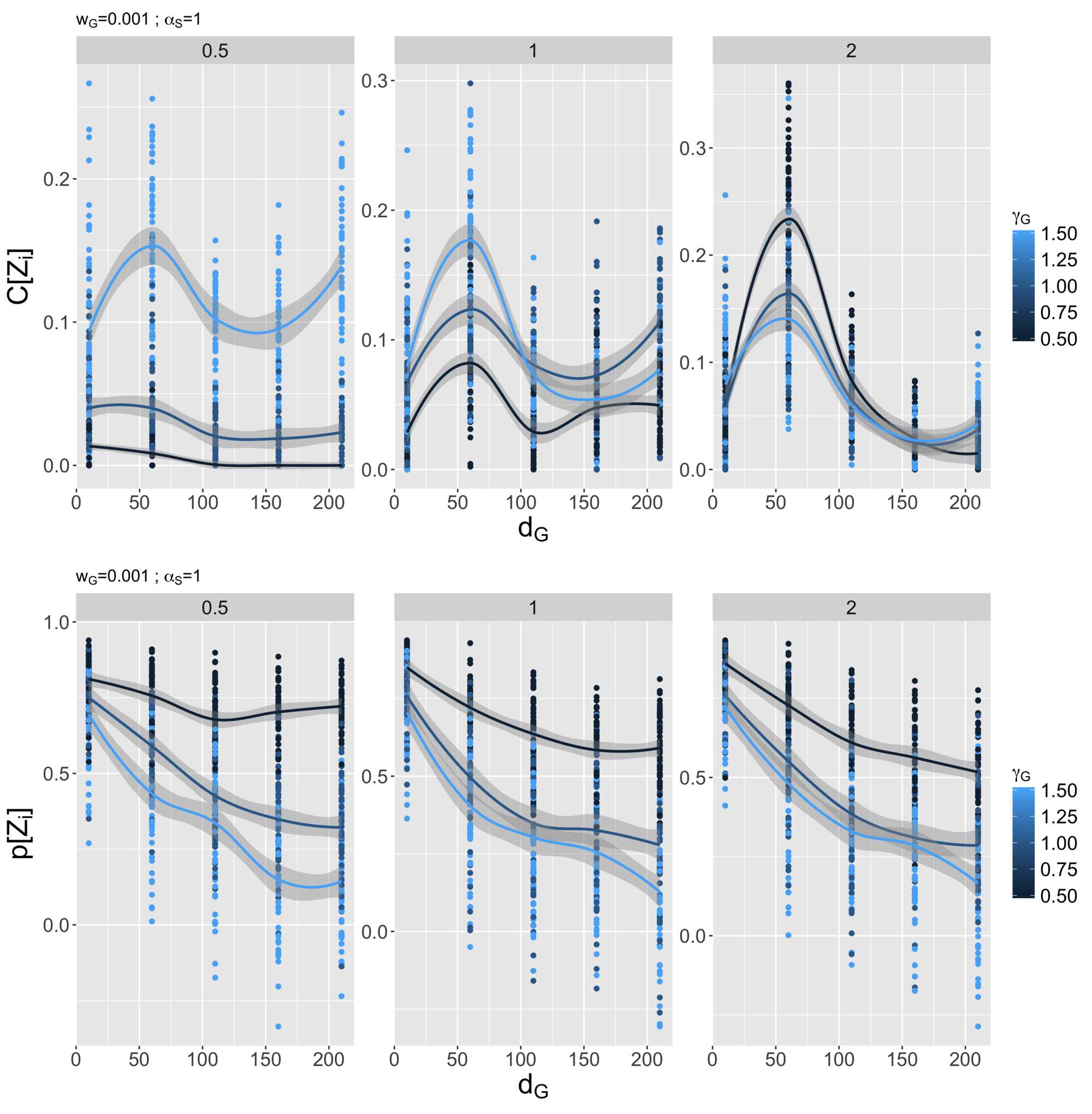}
\caption{\textbf{Agregated behavior of the co-evolution model.} \textit{(Top)} Complexity of accessibilities, as a function of $d_G$, for $\phi_0$ (columns) and $\gamma_G$ (color) variable, at fixed $w_G = 0.001$; \textit{(Bottom)} Rank correlations of accessibilities as a function of $d_G$, for the same parameters.\label{fig:macrocoevol:behavior-aggreg}}
\end{figure}

The evolution of the average closeness centrality in time is shown in Fig.~\ref{fig:macrocoevol:behavior-time} (top) for $w_G = 0.001$, and with variables $(\gamma_G,\phi_0)$ (the behavior is not sensitive to $d_G$). This evolution witnesses a transition as a function of the level of hierarchy: when it decreases, we observe the emergence of trajectories for which the average centrality increases in time, what corresponds to configurations in which all cities profit in average from accessibility gains. This transition can be understood in the sense of \cite{sanders2017peupler}, i.e. a change of the dynamical regime of the urban system, as the qualititative behavior of indicators changes when parameters change.

Concerning the entropy of populations, for which the temporal trajectory is shown in Fig.~\ref{fig:macrocoevol:behavior-time} (bottom), all parameters give a decreasing entropy, i.e. a behavior of convergence of cities trajectories in time. This could be analog to the phenomenon of economic convergence \citep{sachs1995economic}, but at the scale of cities and for populations.

Looking at the complexity of accessibility trajectories, we observe for values of $\phi_0 > 1.5$ a maximum of complexity as a function of interaction distance $d_G$, stable when $w_G$ and $\gamma_G$ vary. This intermediate scale can be interpreted as producing regional subsystems, large enough for each to develop a certain level of complexity, et isolated enough to avoid the convergence of trajectories over the whole system. We reconstruct therein a spatial non-stationarity, and rejoin the concept of the ecological niche (a rather independent ecosystem in which there is co-evolution between the species \citep{holland2012signals}) localized in space: the emergent subsystems that are relatively independent, are good candidates to contain processes of co-evolution. The emergence of this intermediate scale can be compared to the modularity of the French urban system showed by~\cite{berroir2017systemes}. This is to the best of our knowledge the first time that this niche analogy has been identified in a model of a system of cities, although similar conceptualizations in neighbor fields exist such as in political science \citep{monstadt2009conceptualizing} or in the study of technological change \citep{geels2005processes} in which the concept of technological niche is central.

Finally, the behavior of rank correlations for accessibility reveals that the interaction distance systematically increases the number of hierarchy inversions (in the sense of a switch between two cities in their population ranks), what corresponds in a sense to an increase in overall system complexity. The hierarchy parameter diminishes this correlation, what means that a more hierarchical organization will impact a larger number of cities in the qualitative aspects of their trajectories. This effect is similar to the ``first mover advantage'' showed by \cite{levinson2011does}, which unveils a path dependency and an advantage to be rapidly connected to the network: in our case, the modifications in the hierarchy correspond to cities that benefit from their positioning in the network.

\subsubsection{Co-evolutive behavior}


We can now study the ability of the model to effectively produce co-evolutive dynamics. This property of the model is crucial for several reasons: (i) being based on a strong coupling at the process level (or microscopic level) does not imply that effective statistical patterns of co-evolution (in the sense of the characterization proposed by \cite{raimbault2018caracterisation}) emerge from model behavior at the macroscopic level; (ii) there is indeed no example in the literature in which such a link was made; (iii) the spectrum of dynamical regimes the model can produce will inform on the actual existence of structuring effects (direct causality without co-evolution) or of more intricate relation in the case of co-evolution.

To characterize co-evolutive dynamics, we apply a simple but generic method, particularly suited for the study of spatio-temporal causality patterns in territorial systems, which was recently introduced by \cite{raimbault2017identification}. This method uses a weak version of Granger causality, by classifying profiles of lagged correlations between variables. If an absolute maximum of correlations for a strictly positive or negative lag exists, then we have a direct causal link between the variables. A co-evolution between two variables $X,Y$ in that sense will correspond to the existence of circular causal links $X\rightarrow Y$ and $Y\rightarrow X$. We denote here by $\rho_{\tau}\left[X,Y\right]$ the lagged correlation with delay $\tau$.

The exploration of profiles for $\rho_\tau$ for varying parameter values suggests the existence of multiple causality regimes. We however observe (i) the systematic existence of a constant correlation at $\tau = 0$ and (ii) the small variations of correlations that impose the need for a statistical test to ensure that we isolate a significant effect. Therefore, we extend the original method and impose here an additional statistical test: for $\tau_+ = \textrm{argmax}_{\tau>0} \left|\rho_{\tau} - \rho_0\right|$ and $\tau_- = \textrm{argmax}_{\tau<0} \left|\rho_{\tau} - \rho_0\right|$, a Kolmogorov-Smirnov test is used to compare the distributions of $\rho_{\tau_{\pm}}$ and of $\rho_0$. If they are declared different with a p-value smaller than $0.01$, and if $\left|\rho_{\tau_{\pm}}\right| > \left|\rho_0\right|$, we accept the causality link between variables in the corresponding direction. A configuration is then coded by a representation of its graph between variables, given by the six discrete variables equal to 0 if there is no link between the variables (within all directed couples between population, accessibility and centrality) and 1 or -1 depending on the sign of the correlation if there exists a statistically significant link (in practice we observe only positive correlations).


\begin{figure}
	\includegraphics[width=\linewidth]{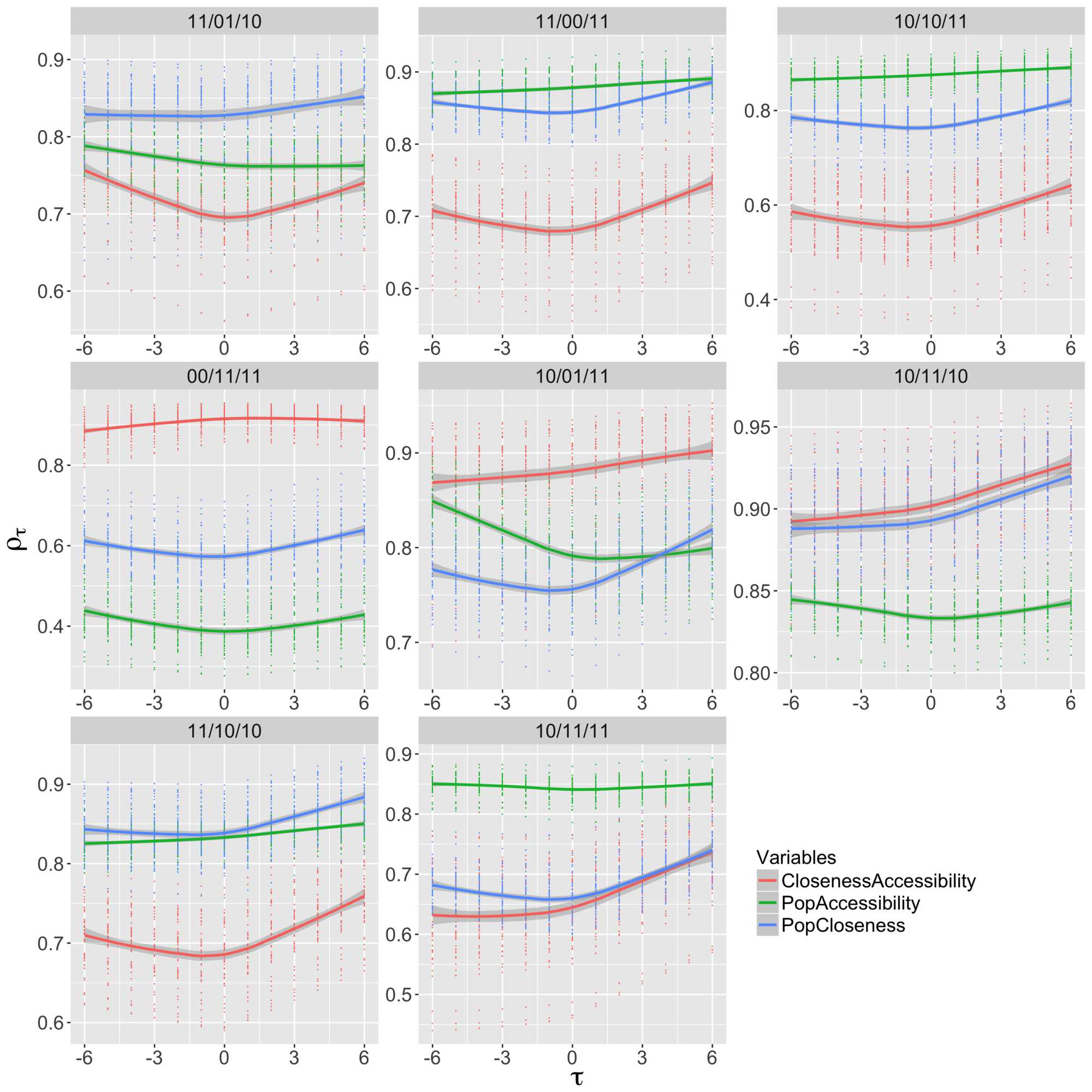}
\caption{\textbf{Lagged correlations.} We give here for the 8 configurations showing at least 4 links between variables (coded in the order of couples, by the existence or not of a link for $\tau_+$ and for $\tau_-$), the lagged correlation profiles $\rho_{\tau}$ as a function of $\tau$, for all couples of variables (color).\label{fig:macrocoevol:correlations}}
\end{figure}

We obtain overall 33 different configurations of links between variables, out of the 64 possible configurations ($2^6$ possible choices for positive correlations only). The type of relations we obtain are particularly interesting regarding co-evolution. We indeed observe:
\begin{itemize}
\item a configuration without any link between variables;
\item 13 configurations of type ``structuring effect'', i.e. for which the graph does not have any loop;
\item a configuration of type ``indirect co-evolution'', for which the graph has a loop of length three ($c_i \rightarrow X_i \rightarrow \mu_i \rightarrow c_i$) ;
\item 18 configurations of type ``co-evolution'', in which there exists at least a loop of length two (direct circular relation between two variables).
\end{itemize}

Among all these regimes, 8 correspond to a graph with at least 4 links (which are then necessarily co-evolutive): we show these profiles in Fig.~\ref{fig:macrocoevol:correlations}. Two regimes witness a positive deviation of the correlation between population and accessibility for positive delays, increasing up to the maximal delay, what could be a clue of a reinforcement of population dynamics through centrality, stylized fact shown for the French system of cities by~\cite{bretagnolle:tel-00459720}.

The regimes in which the centrality is co-evolving with population correspond to the ones where the co-evolution between the network and the territory is the strongest (since the accessibility depends on both), and are observed for large values of $d_G$ (average $d_G=183$ on 62 parameter points). This way, this co-evolution is favored by long interaction ranges.

Finally, the regime with the largest number of links (which corresponds to the regime coded by ``10/11/11'', with co-evolution of population and centrality and of population and accessibility, and a causality of centrality on accessibility), is obtained for a long interaction range $d_G = 160$, a strong interaction hierarchy $\gamma_G = 1.5$, but a low hierarchy of the initial system of cities $\alpha_S$: far-reaching but hierarchical interactions in an uniform system of cities lead to a maximum of entanglement between variables.

We finally confirm these results of variety in causality regimes produced by the model by applying the \emph{Pattern Space Exploration} algorithm~\citep{10.1371/journal.pone.0138212} to the model, with objectives the six correlations studied above (evaluated as zero in the case of a non-significance). We mainly obtain a number of regimes produced by the model larger than the ones obtained before (with negative correlations, 260 realized regimes out of $3^6 = 729$ possible). This short complementary study confirms the ability of the model to produce a large number of co-evolution regimes.

\subsection{Application to the French system of cities}

The model is then applied to the French system of cities on long time dynamical data: the Pumain-INED database for populations, spanning from 1831 to 1999 \citep{pumain1986fichier}, with the evolving railway network from 1840 to 2000 \citep{thevenin2013mapping}. Such a time span can be associated with structural effect on long time. This application aims on the one hand at testing the ability of the model to reproduce a real dynamic of co-evolution, and on the other hand at extracting thematic information on processes through calibrated parameter values.

\subsubsection{Data}

We work on railway network data constructed by~\cite{thevenin2013mapping}. The French railway network is particularly interesting jointly with population data already presented, since the covered time span is relatively close, and this transportation mode has at any times materialized the implication of public and private actors. It corresponds to different processes depending on the period, from a more decentralized management to a more centralized recently, and different technological materializations with for example the recent emergence of high speed trains \citep{zembri1997fondements}. For each date in the population database, we extract the simplified abstract network in which all stations and intersections with a degree larger than two are linked with abstract links which speed and length attributes correspond to real values, at a granularity of 1km. This yields the time-distance matrices between the cities included in the model.


\subsubsection{Stylized facts}

\begin{figure}
	\includegraphics[width=\linewidth]{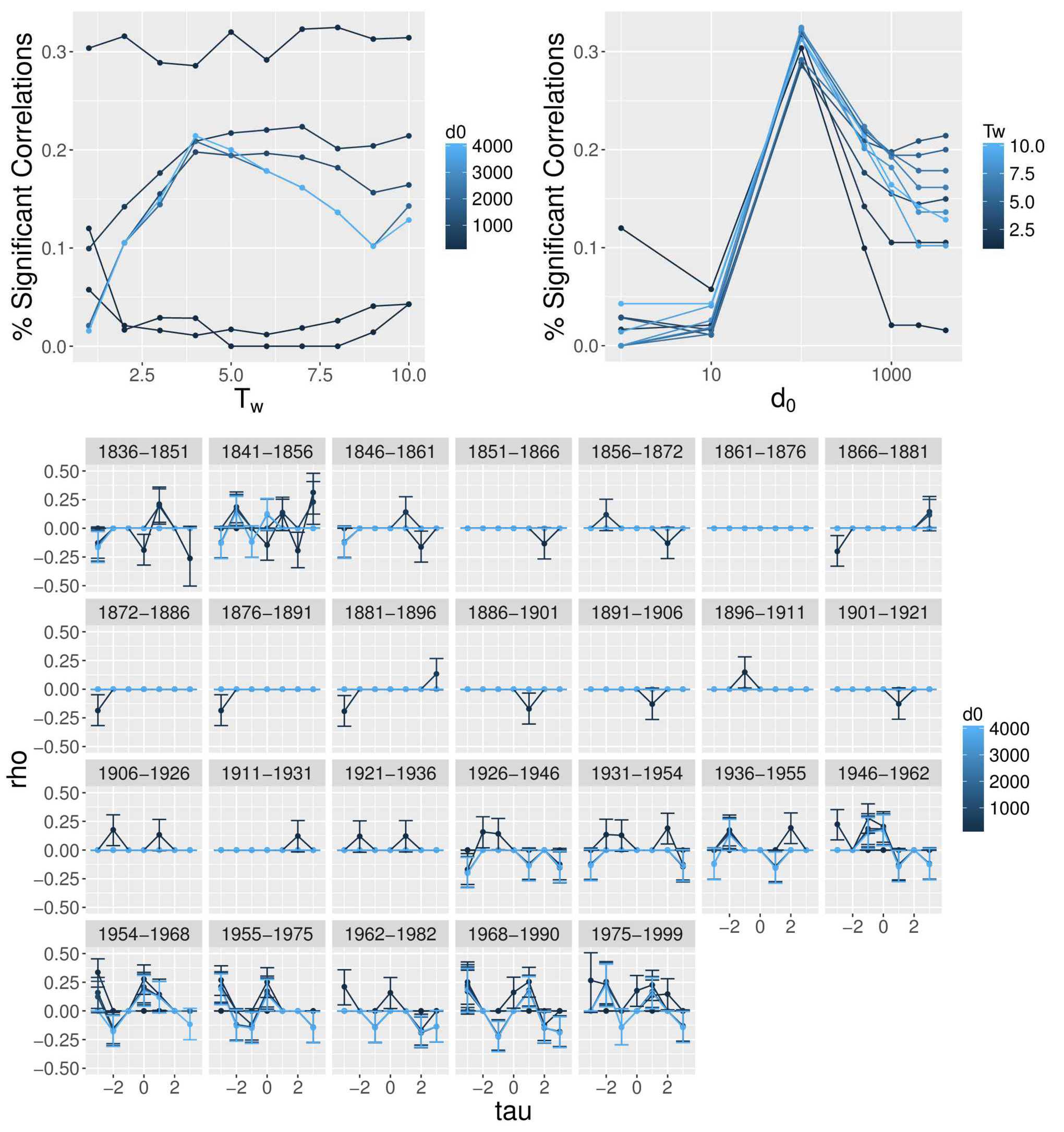}
	\caption[Empirical lagged correlations for the French system of cities]{\textbf{Empirical lagged correlations for the French system of cities.} Correlations are estimated on a window of duration $5\cdot T_w$, between population growth rates and the variations of closeness centrality with a decay parameter $d_0$ (see text). \textit{(Top left)} Number of significant correlations (taken such that $p<0.1$ at 95\%) as a function of $T_w$ for $d_0$ variable; \textit{(Top right)} Number of significant correlations as a function of $d_0$ for $T_w$ variable; \textit{(Bottom)} For the ``optimal'' window $T_w = 4$, value of $\rho_{\tau}$ as a function of $\tau$, for all successive periods.\label{fig:macrocoevol:empirical}}
\end{figure}

Before calibrating the model, we can observe the lagged correlation patterns in the dataset, by applying the causality regimes method \citep{raimbault2017identification}. This empirical study should on the one hand allow us to verify well known stylized facts, and on the other hand to produce a preliminary knowledge of empirical system behavior. We compute as detailed above the closeness centrality through the network, given by $T_i = \sum_j \exp{-d_{ij}/d_0}$, and we study the lagged correlation between its derivative $\Delta T_i$ and the derivative of the population $\Delta P_i$, given by $\hat{\rho}_{\tau} = \hat{\rho}\left[\Delta P_i(t),\Delta T_i(t-\tau)\right]$ estimated on a moving window containing $T_w$ successive dates. We show in Fig.~\ref{fig:macrocoevol:empirical} the results obtained.

These results are important for at least two reasons. First, the behavior of the number of significant correlations as a function of $T_w$ and $d_0$ allows us to find stationarity scales in the system. We observe on the one hand a specific spatial scale that gives a maximum for all temporal windows, at $d_0 = 100km$, what suggests the existence of consistent regional subsystems, which existence is stable in time: indeed, this value corresponds to the interaction distance. It remarkably coincides with the intermediate scale isolated in the synthetic model. On the other hand, long spatial ranges induce an optimal temporal scale, for $T_w = 4$ what corresponds to around twenty years: we identify it as the overall temporal stationarity scale of the system and study the lagged correlations for this value.

Secondly, the behavior of lagged correlations does not seem to comply to the existing literature. At the intermediate spatial scale, the values of $\rho_+,\rho_-$ exhibit no regularity. On the whole system, there is until 1946 close to no significant effect, then no causality between 1946 and 1975 (maximum at $\tau = 0$, non-significant minimum), and a 5 years shift of accessibility causing population after 1968 (the effect staying however doubtful). We do not reproduce the correlation effect between network centrality and place in the urban hierarchy advocated by~\cite{bretagnolle2003vitesse}, what lead us to question the existence of the ``structural co-evolution'' on long time described by \noun{Bretagnolle} in~\citep{espacegeo2014effets}. What \cite{bretagnolle2003vitesse} obtains is a simultaneous correspondence between growth rate and level of connectivity to the network (and not with network dynamic), but not in our sense a co-evolution, since no statistical relation is furthermore exhibited.


We rejoin the recent results of~\cite{mimeur:hal-01616746} that show the statistical non-significance of the correlation between growth rate and evolution of network coverage and accessibility, at a zero delay. Our results are less precise on the class of cities studied (they differentiate large and small cities, and work on a larger panel), but more general as they study variable delays and accessibility ranges, and are thus complementary.

\subsubsection{Calibration of the abstract model}

Expected results of the calibration on real data concern both the more or less accurate reproduction of real city population growth dynamics, i.e. to what extent the inclusion of a dynamical network can increase the explanatory power for trajectories, and also how realistic the evolution of network distance is. We still work with the abstract model.

\paragraph{Model evaluation}

The population trajectories are evaluated similarly to \cite{raimbault2018indirect}, with the logarithm of the mean square error on cities and on time for populations. We add a calibration indicator for distance, given by
\[
\varepsilon_D = \log \left[ \sum_t \sum_{i,j} \left(d_{ij}(t) - \tilde{d}_{ij}(t)\right)^2\right]
\]
where $d_{ij}(t)$ are observed distances and $\tilde{d}_{ij}(t)$ the simulated distances. It is simply a cumulated squared-error, as used for the comparison of origin-destination matrices in a similar case of simulation of a transportation network in~\cite{jacobs2016transport}.


\paragraph{Results}

\begin{figure}
	\includegraphics[width=\linewidth]{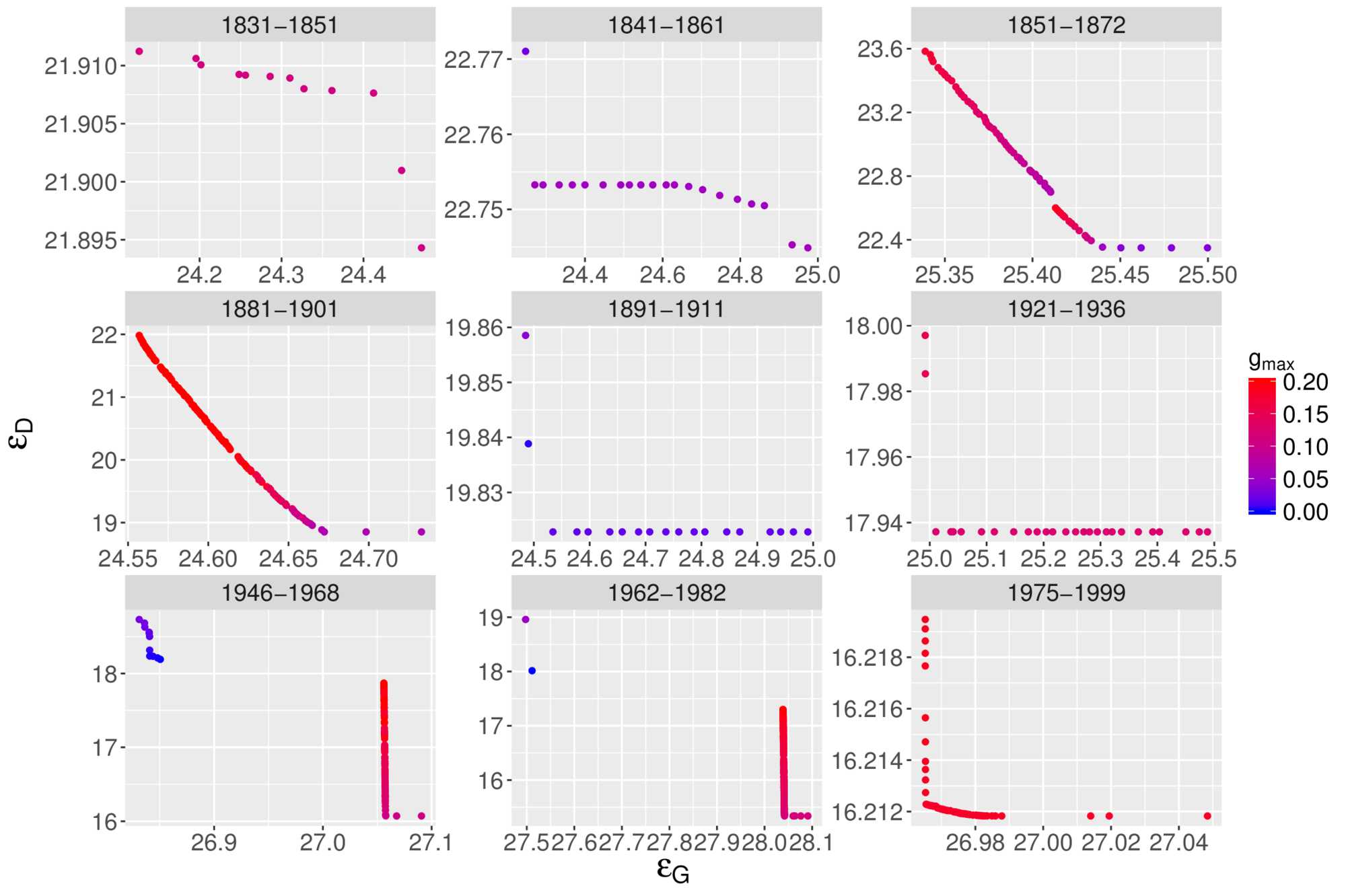}
	\caption[Pareto fronts for the calibration on population and distance]{\textbf{Pareto fronts for the bi-objective calibration between population and distance.} Fronts are given for each calibration period and are colored according to $g_{max}$.\label{fig:macrocoevol:pareto}}
\end{figure}

\begin{figure}
	\includegraphics[width=\linewidth]{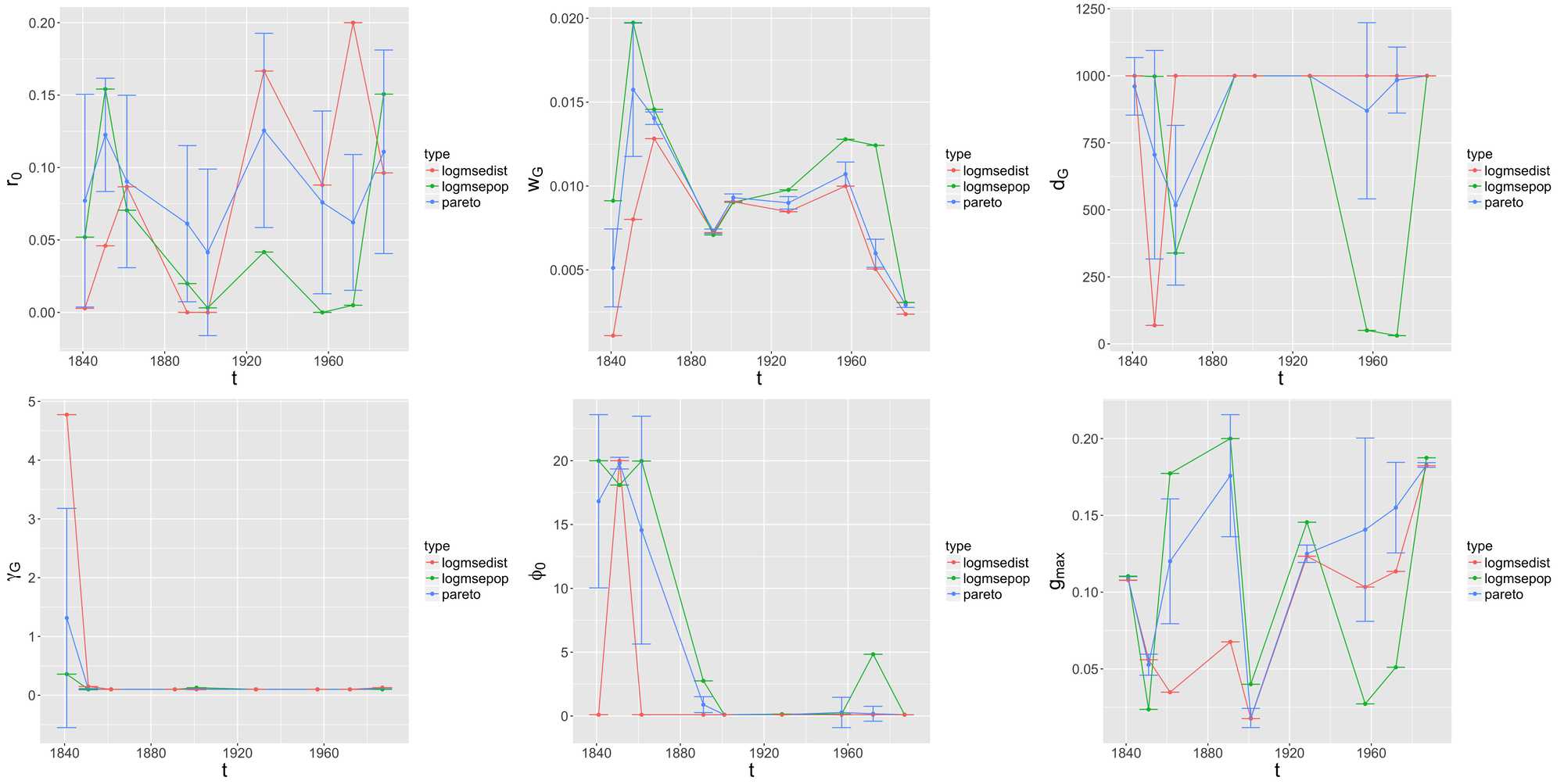}
	\caption[Evolution of calibrated parameters]{\textbf{Temporal evolution of optimal parameters.} From left to right and top to bottom, values of parameters $(r_0,w_G,d_G,\gamma_G,\phi_0,g_{max})$, respectively for the full Pareto front (blue), for the optimal point in the sense of the distance (red) and the optimal point in the sense of the population (green). \label{fig:macrocoevol:parameters}}
\end{figure}

We proceed to a non-stationary calibration, on the $(\varepsilon_P,\varepsilon_D)$ objectives, i.e. the squared-error on populations and on distances. The estimation is done with a moving window on periods with a duration of 20 years. The calibration is done with a standard genetic algorithm (NSGA2) provided by the model exploration platform OpenMole \citep{reuillon2013openmole}. The Fig.~\ref{fig:macrocoevol:pareto} shows the obtained Pareto fronts, and the Fig.~\ref{fig:macrocoevol:parameters} the evolution in time of parameter values for the optimal solutions.

We observe a large variability of the shape of Pareto fronts for the bi-objective calibration on population and distance, what witnesses more or less difficulty to simultaneously adjust population and distance. Some periods, such as 1891-1911 and 1921-1936, are close to have a simultaneous objective point for the two objectives, what would correspond to a good correspondence of the model to both trajectories of cities and trajectory of the network on these periods. 

In comparison with calibration results of the model with static network of \cite{raimbault2018indirect}, when comparing the performances for the objective $\varepsilon_G$, we find periods where the static is clearly better (1831 and 1841 for example) and others where the co-evolutive model is better (1946 and 1962): thus, taking into account the co-evolution helps in some cases to have a better reproduction of population trajectories.

The values of optimal parameters in time, shown in Fig.~\ref{fig:macrocoevol:parameters}, seem to contain some signal. The evolution of $w_G$ and $\gamma_G$ are coherent with the evolutions observed for the static model. For $d_G$, the model principally saturates on the maximal distance and the evolution is difficult to interpret. However, the evolution of $\phi_0$ could be a sign of a ``TGV effect'' in recent periods, through the secondary peak for population after 1960. Indeed, the construction of high speed lines has shortened distances between cities on top of the hierarchy, and an increase of the threshold $\phi_0$ corresponds to an increase of the selectivity for a potential diminution of distances.

The calibrated $g_{max}$ can finally be interpreted according to the history of the railway network (at least of all points in the Pareto front): a significant secondary peak in the first years, a minimum in the years corresponding to the stabilization of the network (1900), and an increase until today linked to the increase of train speeds and the opening of high speed lines.

We have this way in a certain extent indirectly quantify interaction processes through the network and the processes of network adaptation to flows, in the case of a real system.

\subsubsection{Model with a physical network}

We finally sketch the outline of a specification of the model with a physical network, what would in a sense correspond to an hybrid model combining different scales. The objective of such a specification would be on the one hand to study the difference in trajectories compared to the abstract network, i.e. to quantify the importance of economies of scale (due to common links), of congestion and also the possible compromises to take in order to spatialize the network. On the other hand, it would help to understand to what extent it is possible to produce realistic networks in comparison to autonomous network growth models (see \cite{xie2009modeling}) for example. This specification follows the frame of \cite{li2014modeling}, which model the co-evolution between transportation corridors and the growth of main poles at a regional scale. Note that possible ontologies for fully mesoscopic co-evolution models seem to be very different from the one we used here, closer to cellular automata urban morphogenesis models \citep{2018arXiv180505195R}. They can in particular include multiple processes for the growth of transportation infrastructure networks \citep{raimbault2018multi}.

The physical network we implement aims at satisfying a greedy criteria of local time gain. More precisely, we assume a self-reinforcement similar to~\cite{tero2010rules} A specification analog to the one used before assumes a growth for each link, given also in a logic of self-reinforcement by:

\[
d(t+1) = d(t)\cdot \left(1 + g_{max} \cdot \left[\frac{\phi}{\max \phi}\right]^{\gamma_s}\right)
\]

if $\phi$ is the flow in the link and $d(t)$ its effective distance. The threshold specification used before does indeed not allow a good convergence in time, in particular with the emergence of local oscillation phenomena.

\begin{figure}
	\includegraphics[width=\linewidth]{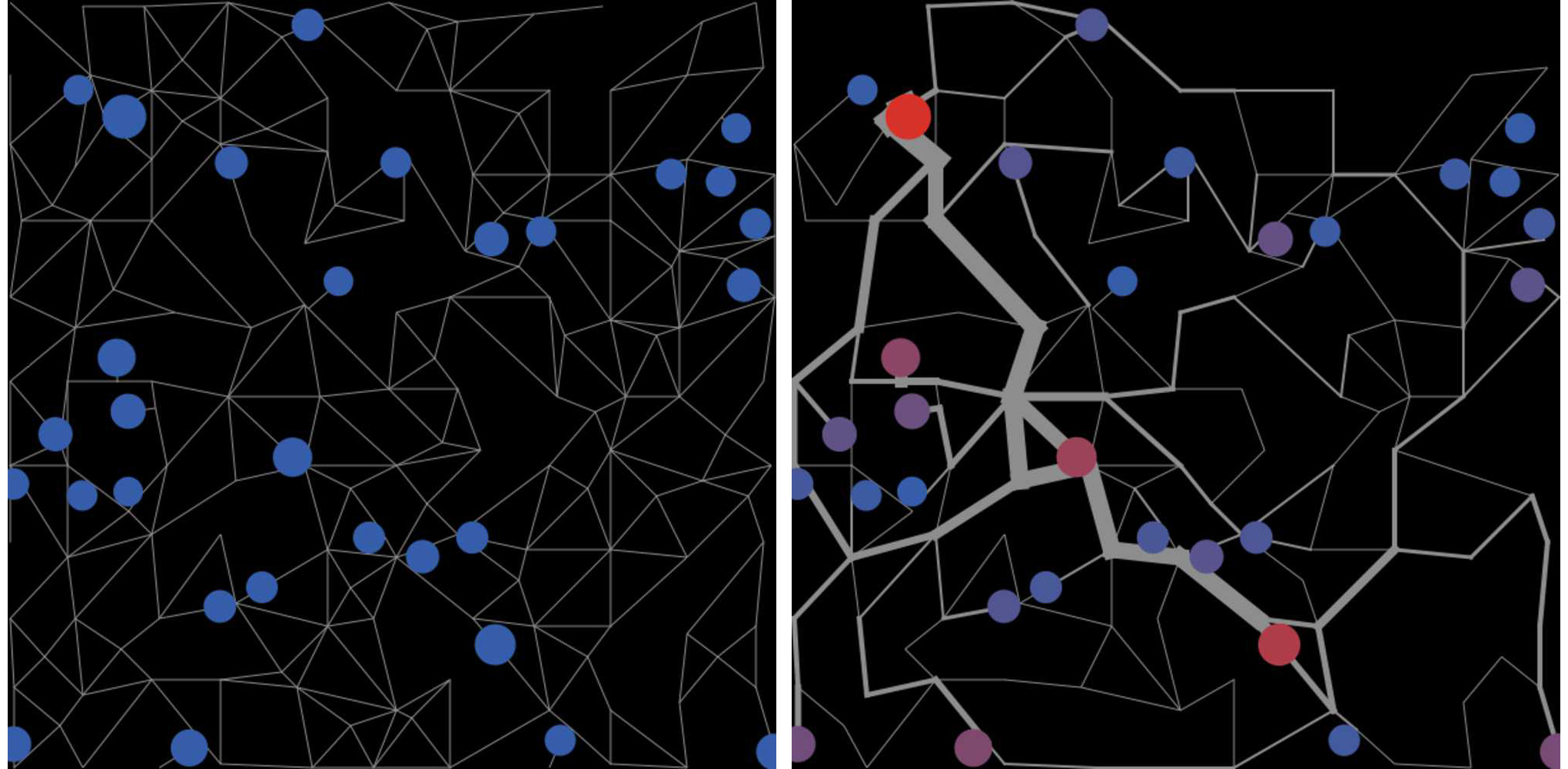}
	\caption[Example of application of the macroscopic model with a self-reinforcing network]{\textbf{Example of configuration obtained with a self-reinforcing network.} \textit{(Left)} Inital random configuration, with uniform impedances; \textit{(Right)} Final configuration obtained after 100 iterations. Circles are cities distributed in space, which relative size gives the population and color level the total population growth. Physical transportation links are in grey and their thickness gives the flow traversing them.\label{fig:macrocoevolution:slimemould}}
\end{figure}

We generate a random initial network, by perturbing the position of vertices of a grid for which a fixed proportion of links has been removed (40\%) and by linking cities to the network through the shortest path. Links have all the same impedance, which then evolves according to the equation above. An example of a configuration obtained with this specification is given in Fig.~\ref{fig:macrocoevolution:slimemould}. The good convergence properties (visual stabilization of network structure during restricted experiments) suggest the potentialities offered by this specification, which systematic exploration is out of the scope of this work.

\section{Discussion}

The study of particular trajectories within a system of cities can allow to answer to specific thematic questions: for example, the influence of medium-sized cities on the global trajectory of the system can be assessed through this model. More precisely, depending on parameters, a given city can more or less influence the trajectories of all cities and the network evolution. Very high hierarchy parameter will favor only the biggest cities, but lower values will give more importance to the rest of the distribution, in particular medium-sized cities. The quantification of their contribution in indicator variance could for example inform on their role in the system. The drivers of a more or less ``successful'' trajectory for such medium-sized cities can also be investigated, i.e. if a higher population growth or connectivity is obtained thanks to a higher centrality or thanks to the proximity of a large city for example. In the case of the application to a real system, the mapping of deviation to the model in time can suggest regional particularities, i.e. local factors exogenous to the model that have a high impact on population growth patterns.

We also finally expect to be able through the model to compare urban systems in different geographical and political contexts, and at different scales. This should foster the understanding the implications of planning actions on the interactions between networks and territories. For example, French railway network has emerged through multiple operators, on the contrary to the Chinese high speed railway network, for which a more precise development could be considered.

An other development would be to superpose different layers of the transportation network as different multi-modeling components, with for example a dynamical highway database for the case study we took. This would allow to make the model more complex and compare which transportation mode fits better within co-evolutionary dynamics. In a broader perspective, the family of models we introduced could be used for a more general benchmark of models of growth for systems of cities based on interactions, that would also include for example the Favaro-Pumain model focusing on innovation \citep{favaro2011gibrat}, or the Marius model family focusing on economic dynamics \citep{cottineau2014evolution}. A systematic comparison of such models on various systems of cities would shed light on concurrent explanations for city growth and possible superposition of processes.

\section*{Conclusion}

We have given a first insight into models of co-evolution for cities and networks at the macroscopic scales, by introducing a family of models and studying the properties of a specification on synthetic data and on the French system of cities. This work paves the way for a finer understanding of entangled interactions in systems of cities.

Going back to the question of structuring effects of transportation networks, we have confirmed from the modeling and empirical perspective that the issue is highly complex: (i) our empirical analysis for the French system of cities does not reveal any possible effect, in contradiction with previous results in the literature; (ii) our modeling experiments show that there exist theoretically some dynamical regimes in which these effects actually exist, and others in which interactions are more intricate, in particular when there is co-evolution. The empirical identification of these on real systems, e.g. through a correspondance between model regimes and calibrated parameters, remains an open issue out of the scope of this work, for example due to issues of equifinality for which efficient inverse problem heuristic for simulation models must be established.

\section*{Acknowledgements}

Results obtained in this paper were computed on the vo.complex-system.eu virtual organization of the European Grid Infrastructure ( http://www.egi.eu ). We thank the European Grid Infrastructure and its supporting National Grid Initiatives (France-Grilles in particular) for providing the technical support and infrastructure.

\end{document}